\documentclass[twocolumn,prl,nofootinbib]{revtex4-1}
\usepackage{amsmath,amssymb}
\usepackage{verbatim,graphicx}
\usepackage{color}

\def\Tr{\textrm{Tr}}
\def\lp{\left(}
\def\rp{\right)}

\begin{document}
\setlength{\unitlength}{1mm}
\title{Numerical Evidence for Firewalls}

\author{David Berenstein and Eric Dzienkowski}
\affiliation{Department of Physics, University of California at Santa Barbara, CA 93106}

\begin{abstract}
We study thermal configurations in the BFSS matrix model produced by numerical simulations. 
We do this by adding a probe brane to a typical configuration and studying the fermionic degrees of freedom connecting the probe to the configuration. 
Depending on the parameters of the probe there is a region where these fermion modes are gapless. 
We argue that it is natural to excise the gapless region from the geometry and that the black hole horizon is located exactly at the edge of this zone.
The physics inside the gapless region is maximally non-local and effectively 1+1 dimensional.
When a probe, considered as an observer, crosses into the gapless zone there is a break down of effective field theory where the off-diagonal fermions are integrated out.
We argue that this breakdown of effective field theory is evidence for firewalls on black hole horizons.
\end{abstract}

\maketitle

\begin{center}{\em Introduction}
\label{sec:Introduction}
\end{center}

Hawking's discovery that quantum black holes radiate and behave as thermal systems \cite{Hawking:1974sw} has led to the black hole information paradox. 
Recent work on this problem with a modern understanding of quantum information theory \cite{Braunstein:2009my,Almheiri:2012rt,Almheiri:2013hfa} (see also \cite{Marolf:2013dba,Bousso:2013wia}) has suggested that black hole horizons (at the very least for old black holes) are not smooth right behind the horizon.
The geometry is instead replaced by a so called ``firewall" where an infalling observer would be destroyed on contact, before reaching the classical black hole singularity.
The postulates that lead to this statement are:
	{\em i)} black hole formation and evaporation processes can be addressed entirely within the context of quantum mechanics,
	{\em ii)} outside the black hole, physics can be addressed semiclassically,
	{\em iii)} to an outside observer the black hole appears as a system with discrete energy levels reproducing the entropy of the black hole and finally
	{\em iv)} an infalling observer experiences no drama when crossing the horizon.
These four postulates together were shown to be inconsistent.
The gauge/gravity duality gives an example of the first three postulates \cite{Maldacena:1997re} being correct.
Thus, the fourth postulate about infalling observers seems to be the one that needs to be removed.
A rather important question is then what happens to an infalling observer: can we actually see what kind of drama it would experience?

We will address this question for a special class of black holes subject to the gauge/gravity duality.
Using the gauge theory as a microscopic theory of gravity, all questions will be addressed within the dynamics of the gauge theory. 
We study black holes in the BFSS matrix model \cite{Banks:1996vh}.
Our treatment follows the numerical classical simulations carried out in \cite{Asplund:2011qj,Asplund:2012tg}.
The black holes in question are {\em hot and stringy}. 
It is not obvious that our conclusions carry over to {\em cold} semiclassical black holes.
We will give arguments for the robustness of our reasoning. 

The main problem we need to address is whether we can find a geometric characterization of the black hole horizon in the matrix model.
Without such a characterization, we can not ask what happens when we cross the horizon; we can not even tell where the horizon is located to ask such question.
We find that the ideas of previous work \cite{Berenstein:2012ts} extended to the BFSS matrix model give a possible characterization and a geometric separation between the inside and outside of the black hole.
The main idea is to consider a point-like probe brane (that is, a D0-brane) in the presence of the black hole.
The probe brane acts as an observer which explores an $\mathbb{R}^9$ geometry.
We can ask where and under which conditions one can integrate out the degrees of freedom connecting the probe to the black hole.
In our setup we care mostly about the fermionic degrees of freedom. The same problem has been studied in \cite{Iizuka:2001cw} using mean filed theory methods where the probe is fully dynamical, but outside the black hole. 
When these degrees of freedom can not be integrated out, the physics of the observer changes.
We will see that there is a region of ${\mathbb R}^9$ where the fermionic spectrum of modes connecting the probe to the black hole is gapless.
We will also show that the physics in this region can be characterized as very non-local and in a technical sense, the spacetime is $1+1$ dimensional.
It is natural to excise such a very non-local region from the geometry in gravity.
A natural place for a horizon is exactly at the boundary of where the gapless region appears.
We argue that the change in effective physics is {\em very dramatic} and the observer can not fail to notice that the environment has changed completely.
This suggests a precise description for what a firewall is.

\begin{center}
{\em Fermionic Observables and a Gapless Region}
\end{center}
\label{sec:gapless}

A typical problem in matrix theory is to describe a configuration of matrices geometrically. 
The generic solution for a set of commuting matrices is a set of points which characterize the eigenvalues of the matrices \cite{Banks:1996vh}.
The goal of \cite{Berenstein:2012ts} was to give a geometric prescription to configurations of three Hermitian matrices which do not necessarily commute.
The prescription is to measure the spectrum of fermions connecting a D0-brane probe to a background matrix configuration.
The energy spectrum of fermions gives a sense of distance because for strings, their length is proportional to their energy. 
The minimal length string, corresponding to the minimal energy mode, leads to an effective notion of distance.
These ideas have a straightforward generalization to any dimension.

To apply these ideas, we need the fermionic part of the BFSS Hamiltonian \cite{Banks:1996vh}
\begin{equation}
H_{\text{ferm}} \sim \Tr\lp\Psi^\dag\Gamma^i[X^i,\Psi]\rp
\end{equation}
where $1\leq i\leq 9$, the $\Psi$ are $SO(9)$ spinors, and $\Gamma^i$ the nine dimensional gamma matrices.
The constant of proportionality depends on the normalization of the fields and $\hbar$ (in the BFSS matrix model we can trade $g_{YM}$ for $\hbar$ by rescaling the fields).
The matrices $X^i$ and $\Psi$ are in the adjoint of $U(N)$, that is, they are Hermitian.
Given an $N\times N$ configuration of $X^i$ we construct an $N+1\times N+1$ configuration $\tilde{X}^i$ by adding a single D0-brane probe with coordinates $x^i\in\mathbb{R}^9$ in the lowest right corner.
We only want to consider the fermion modes connecting the background and the brane probe.
\begin{equation}
\tilde{X}^i = \begin{pmatrix} X^i & 0 \\ 0 & x^i \end{pmatrix},\quad \tilde{\Psi} = \begin{pmatrix} 0 & \psi \\ 0 & 0 \end{pmatrix}
\end{equation}
where $\psi$ is an $N\times 1$ column vector (it is in the fundamental of $U(N)$).
The effective Hamiltonian for the configuration $(\tilde{X}^i,\tilde{\Psi})$ is 
\begin{equation}
\label{eq:heff}
H_{\text{eff}} \sim \psi^\dag([X^i - x^iI_N]\otimes \Gamma^i)\psi 
\end{equation}
This effective Hamiltonian can be considered at each instant of time and for each position of the probe.
Furthermore, we can choose to make the probe fully dynamical or not. 
If we make it dynamical, we can call it an `observer' and choose a set of initial conditions, that is, the initial position and velocity of the probe.
If the probe is not dynamical, we can scan over $\mathbb{R}^9$ with the position of the probe and label each point by the properties of the fermion spectrum of eigenvalues in equation \eqref{eq:heff}.

\begin{figure}[h]
\includegraphics[width=7cm]{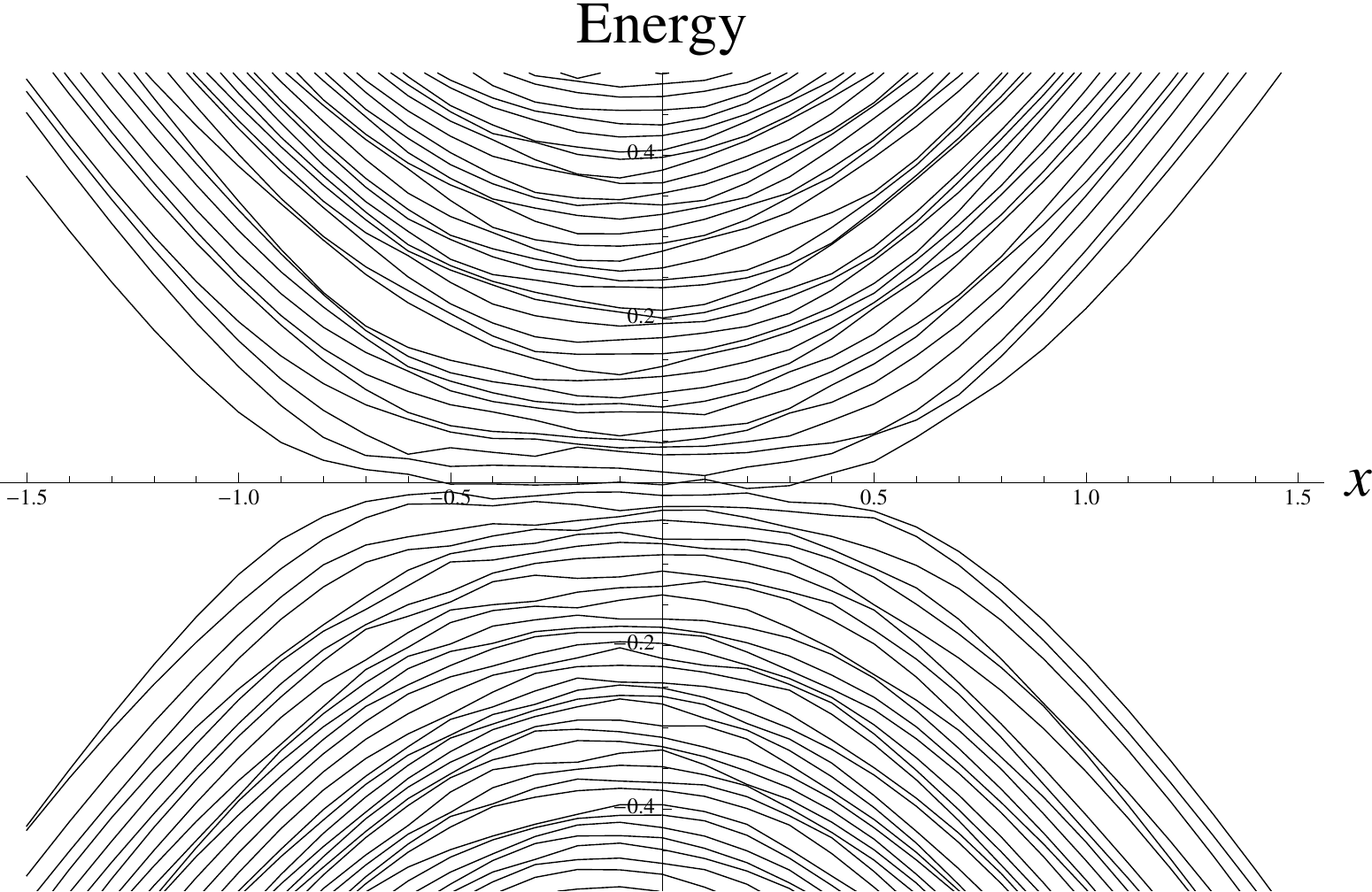}
\caption{Fermionic eigenvalue spectrum for a typical configuration of matrices with $N=47$ after thermalization. The units are arbitrary.}
\label{fig:eigenvalues}
\end{figure}

We consider the spectrum of $H_{\text{eff}}$ for the thermal configurations of \cite{Asplund:2012tg}.
Without loss of generality we may take the matrices to be traceless.
By rotational invariance (the configurations do not have angular momentum), we need only consider moving the brane probe in a single direction, say $x^1$, and we choose to go through the center of the configuration.
The spectrum of $H_{\text{eff}}$ for a typical configuration is plotted in FIG. \ref{fig:eigenvalues}.
Since the nine dimensional gamma matrices are $16\times16$ the spectrum of fermions for a rank $N$ configuration has $16N$ modes.
The spectrum is typically non-degenerate and there are crossings of zero: a fermion becomes massless at such loci. 
Counting such crossings gives a notion of a local index \cite{Berenstein:2012ts}, and this counts the number of strings created when the probe is moved from infinity to a place, giving a generalized version of the Hanany-Witten process \cite{Hanany:1996ie}.

The figure shows a clean separation of two regions.
In the first region, the eigenvalues of the fermions are well separated from zero: there is a gap which can be used to measure the distance from the probe to the configuration.
There is a second region where the typical eigenvalue separation of the fermion spectrum from zero is similar to the separation of the eigenvalues from each other. 
This region is said to be gapless.
Remember that the simulated matrix configurations are thermal.
The temperature gives a thermal activation energy $kT$. 
So long as the modes near zero have energies below $kT$, they can become thermally active. 
At large $N$ and fixed temperature, the size of the matrix configuration $X^i$ scales as $N^{1/4}$ (the typical eigenvalue of $X^i$ scales as $N^{1/4}T^{1/4}$ \cite{Asplund:2012tg}) and so the number of fermions in the band of energy $kT$ grows as $N^{3/4}\simeq N N^{-1/4}$.
Thus, in a large $N$ setup there are a lot of fermion states that could in principle get activated.
In this sense, there is no gap.

\begin{center}
{\em Spectral Dimension and Nonlocality}\label{sec:spectral}
\end{center}

We wish to understand the nature of the physics inside the gapless region, and in particular if we can understand this information geometrically or not.
One step in this direction is knowing the effective dimension of spacetime that the matrix configuration describes.
Since we are exploring $\mathbb{R}^9$, it is natural to suspect that the matrix configuration in the gapless region is $9+1$ dimensional.
In massless free field theories in $d+1$ dimensions, the density of states near zero follows a power law $\rho(\epsilon) \sim \epsilon^{\gamma-1}$ where $\epsilon$ is the energy and $\gamma = d$. 
We argue that if we measure $\gamma$ we can measure the effective dimensionality of the matrix configuration. 
The parameter $\gamma$ will be called the spectral dimension of a configuration.
We argue that if we measure $\gamma$ with a probe we are actually measuring the dimensionality in standard configurations.

Consider a toy model for matrix black holes in which the matrices $X$ are essentially commuting and describe a gas of D0-branes (this has been argued recently for example in \cite{Iizuka:2013yla,Wiseman:2013cda} and references therein). 
The matrices can be diagonalized simultaneously, and we can talk about the positions of the D0-branes as the common eigenvalues in $\mathbb{R}^9$. 
More importantly, we can think of a density of eigenvalues in some region of $\mathbb{R}^9$, call it $\rho(r)$. 
The energy of the fermions connecting a probe at $x$ to an eigenvalue at $r$ will be $\epsilon\simeq |x - r|$.
The number of such states at fixed $|x - r| = s$ for small $s$ is $n(s) = \int d^9r\, \rho(r)\delta(|x - r|- s)\simeq \rho(x) s^8\simeq \rho(x) \epsilon^8$.
So in a region where $\rho(x)\neq 0$ we expect to measure a spectral dimension of $\gamma=9$. 
Similarly, we can consider a D2 brane background obtained from a fuzzy sphere configuration.
If we put the D0-brane probe in contact with the fuzzy sphere, it is easy to show that one gets a spectral dimension $\gamma = 2$, the dimensionality of the sphere as a geometric object. 
This coincides with approximating the fuzzy sphere as a collection of D0 branes uniformly distributed on the surface of the fuzzy sphere.
The spectral dimensionality captures the dimension of extended objects, or equivalently of the smearing of D0 branes in some region.
It is natural to expect that if the matrix configuration can be pictured as some extended D-brane contorted to fill the gapless region, one would measure $\gamma = 9$ just from smearing into a density of D0-branes.
We will now check if this is true or not numerically.
If it does coincide with $\gamma = 9$, we would be giving evidence in favor of the toy model of black holes as a gas of D0-branes or an extended brane filling the region.
We find a completely different result. 

We define
\begin{equation}
\gamma \equiv \lim_{\epsilon\rightarrow 0} \frac{d\ln(\rho(\epsilon))}{d\ln(\epsilon)} + 1
\end{equation}
The difficulty with this definition is that we can not take the limit on a configuration of finite size matrices, where we only have finitely many eigenvalues in the effective Hamiltonian. 
What we need is a fit to a power law by choosing a few points near $\epsilon\simeq 0$. 
The precise way in which we choose to do this can give slightly different answers. 
To reduce such problems, we average over many configurations so that we can measure $\gamma$ statistically.

We also need to compare the spectra of eigenvalues of the effective Hamiltonian at different values of $N$.
This way we can extrapolate to large $N$ and find a value of $\gamma$ that is valid in the thermodynamic limit. 
This helps to show that our result is robust.
To put different values of $N$ on top of each other we take advantage of the scaling symmetry of the classical BFSS matrix model and scale the $X^i$ by some $N$-dependent factor that is also temperature dependent.
Since the $X^i$ approximately follow the Gaussian Unitary Ensemble for traceless Hermitian matrices (TGUE) \cite{Asplund:2012tg}, we can scale the matrices such that their distributions of eigenvalues can be analyzed in terms of the limits of the TGUE which are semicircle distributions.
This is done by fixing the second moment.
The width of the associated semicircle is given by $2\sqrt{N}\sigma$ where $\sigma$ is the width of the TGUE.
The value of $\sigma$ is given by
\begin{equation}
\sigma = \sqrt{\frac{\langle \sum_{i=1}^9 \Tr(X^i) \rangle}{9(N^2 - 1)}}
\end{equation}
where we have averaged over all nine bosonic matrices which is allowed by the $SO(9)$ invariance.

\begin{figure}[h]
\includegraphics[width=7cm]{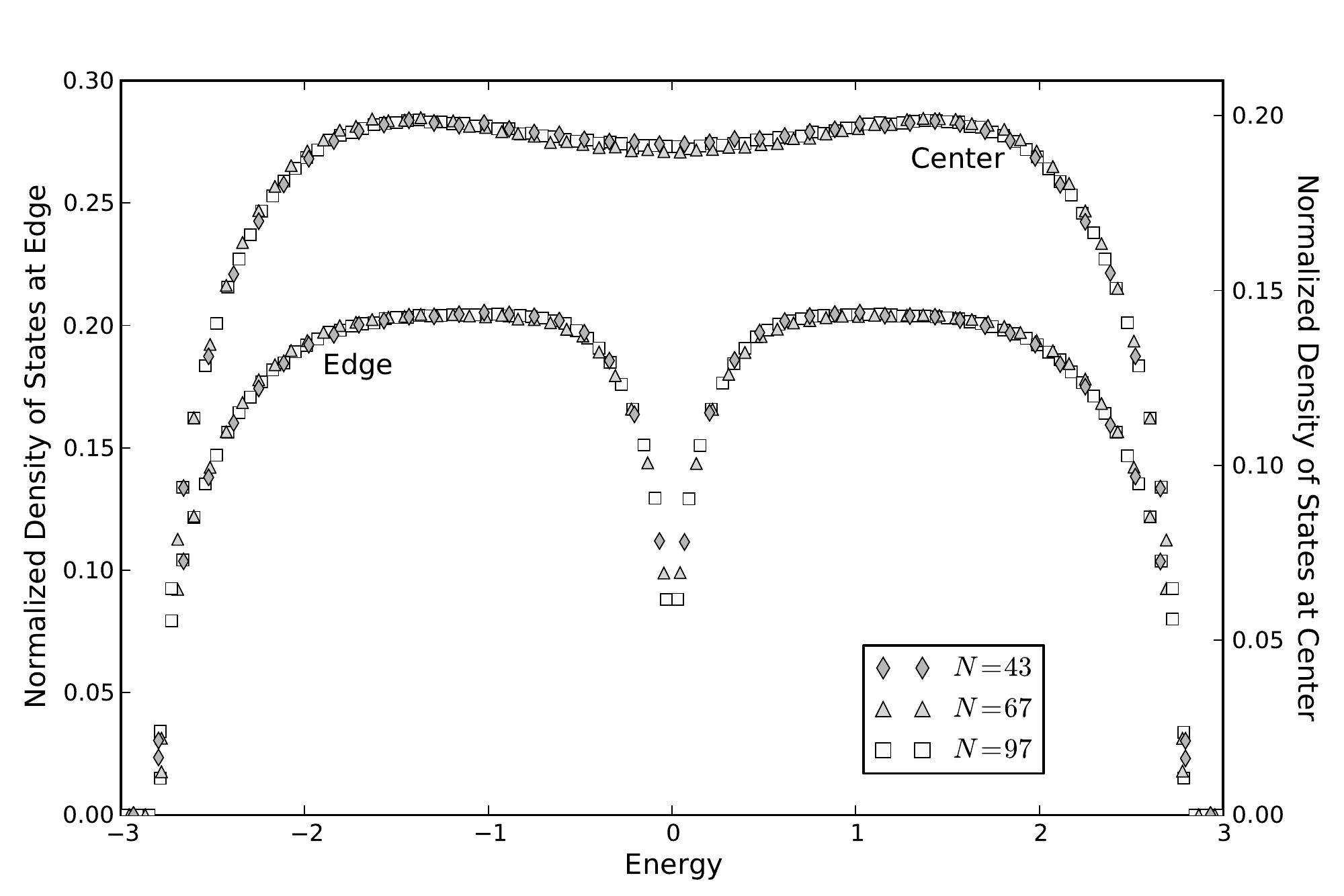}
\caption{A plot of the density of states as a function of energy at the center and edge of the gapless region for various values of $N$ averaged over 1000 configurations.
Each function is normed such that $\int \rho(\epsilon)\, d\epsilon = 1$.}
\label{fig:dos}
\end{figure}

The spectral dimension of the field theory when the probe is at the center of the gapless region is $\gamma = 1.0$. 
This can be seen from FIG. \ref{fig:dos} where the density of states is flat at zero energy. Notice also that the spectral density we have is completely 
different than the one found in \cite{Iizuka:2001cw}. This might be related to the truncations of the dynamics that they did.
As we move away from the center of the gapless region the spectral dimension stays approximately constant.
Near the edge of the gapless region, the spectral dimension shoots up to about $\gamma = 1.3 \pm 0.1$. 
The density of states at the edge shown in FIG. \ref{fig:dos} shows the cusp at zero energy.
It is difficult to pin down exactly where the gapless transition occurs due to a finite bin size and limited statistics, but log-log plots indicate a qualitative change at around $|x| = 0.33 \pm 0.03$ for the rescaled matrix configurations.
For reference, the matrices are scaled so that average maximum eigenvalue approaches one as $N\rightarrow\infty$.
Even with numerical errors the spectral dimension at the boundary is far from nine.
If we include quantum corrections, the value of $\gamma$ should stay close to the classical physics value: in the BFSS matrix model it is believed that there is no finite temperature phase transition between the high temperature regime and the low temperature regime.
This is also verified numerically as the free energy curves as a function of the temperature are smooth \cite{Anagnostopoulos:2007fw,Catterall:2008yz,Hanada:2008ez}. 
Also, the thermal states in the BFSS model are deconfining for arbitrary small temperatures in the dual supergravity setup \cite{Wiseman:2013cda}.

How should we interpret this result?
The authors of \cite{Kaplunovsky:1983ev} consider a class of theories with fermions defined on a fully connected lattice whose links are weighted. 
They argue that the typical configuration is maximally connected giving rise to a theory with one infinite-dimensional symplex which is maximally non-local; every site is connected to every other site by one step.
When the weights are Gaussian distributed, the spectrum of the Hamiltonian follows a cumulative semicircle distribution (in the large $N$ limit).
Near zero energy the spectrum is linear, the density of states is flat, and they conclude that the non-local fermion theory has an effective description in $1+1$ spacetime dimensions, in the sense that the density of states is the same as the one of a $1+1$ dimensional theory.
We also find a spectral dimension of one and thus consider ourselves to be in the same universality class as the models in \cite{Kaplunovsky:1983ev}, which is also the universality class of a random Hamiltonian. 
By analogy to the fully connected lattices, we believe that the probe brane is in an environment where is it effectively equally far from the D0-brane matrix background no matter where the D0-branes are located.
Our numerical results are inconsistent with a local gas of D-branes. 
If this persists in the fully quantum regime, the results of calculations based on models of such gases are suspect.
The setup is maximally non-local, but it looks as if the dynamics lives effectively in $1+1$ dimensions.

\begin{center}{\em Drama at the Horizon}
\label{sec:drama}
\end{center}

We now argue that we should locate the horizon of the black hole exactly at the interface between the gapless region and the gapped region.
This is potentially different than the proposal in \cite{Kabat:1998vc}.
Consider a string suspended between a probe D-brane at fixed position and a black hole. 
The energy carried by such a string is given by $\int_{r_0}^{r_D} T(r) \sqrt{ g_{tt} g_{rr}}\,dr$ where $T(r)$ is the local string tension at $r$, $r_0$ is the horizon, and $r_D$ is the position of the brane. 
This is finite for typical black holes: the black hole horizon is at finite distance. 
In the limit $r_D\to r_0$ we get that the energy of such a string goes to zero.
In our setup, the fermionic energies connecting the probe to the matrix configuration have this same behavior. 
Also, the presence of a large number of massless modes appearing at the edge of the gapless region indicates that wiggles on a string also get redshifted and produce a large number of string states with small energies.
This is analogous to strings spreading when reaching the horizon \cite{Susskind:1993ki}. 

Consider the probe as a dynamical observer moving towards the horizon. 
A natural question to ask is if we can integrate out the modes connecting the probe to the matrix configuration and extract an effective action for the probe.
This is how gravitational interactions and forces between objects are captured in the BFSS matrix model \cite{Banks:1996vh} (see \cite{Taylor:1998tv,Taylor:1999gq} for a systematic treatment). 
In our case the answer to this question is yes, for fermions, away from the gapless region. 

If we reintroduce $\hbar$ and call an eigenvalue of \eqref{eq:heff} $\omega$, the energy of such a fermionic mode is $\hbar|\omega|$. 
The light fermion modes become thermally active when $\hbar|\omega |< kT$. 
In classical physics all the fermions are active, yet we will assume that the system is cold enough so that most fermions are not active.
After taking the limit where $N$ is large with $\hbar$ constant, and rescaling $X$ so that the matrix configuration is of finite size in the probe coordinates, the light fermions become active exactly when we enter the gapless region.
The approximation where we can integrate out the off-diagonal degrees of freedom connecting a probe to the black hole breaks down exactly at the putative horizon.

This breakdown of effective field theory of a probe plus configuration suggests that the observer can not fail to observe that physics has changed dramatically on crossing the horizon  (see however \cite{Horowitz:2009wm}). 
The fourth postulate of \cite{Almheiri:2012rt} was that no drama occurs when crossing the horizon.
Removing this postulate to restore consistency suggested the existence of a firewall.
We see that the BFSS matrix model is giving a physical model for the firewall. 

This in itself does not give a proof that firewalls exist.
There could be some other physical effect on the probe that comes from understanding the bosonic degrees of freedom that forces us to put the horizon elsewhere, as in \cite{Kabat:1998vc}. 
Bosonic instabilities might be realized only via parametric resonance and would require a full time dependent treatment to be understood. 

Our analysis so far has been done for stringy black holes \cite{Asplund:2012tg}. 
For such black holes the horizon and the singularity are on top of each other in string units. 
The $1+1$ dimensional effective physics could be associated with the singularity rather than with the horizon.
This can be remedied with fully quantum simulations \cite{Iizuka:2013yla,Wiseman:2013cda}.

Physics at the black hole singularity is also argued to be effectively $1+1$ dimensional in the causal dynamical triangulation program, since the effective UV structure of gravity has a different dimension \cite{Ambjorn:2005db} (see also \cite{Carlip:2009km}).

Also note that this picture, although similar in spirit to the fuzzball picture \cite{Mathur:2005zp}, is distinct. 
The known fuzzball solutions are geometric (non-singular solutions of supergravity) and they stretch all the way to the horizon.
Microphysics in these setups is essentially gravitational.
In our case the inside of the black hole gets replaced by non-geometric, non-local objects whose effective dimension is different than that of the ambient space. 

The presence of an effective $1+1$ dimensional field theory starting at the horizon is also reminiscent of ideas espoused by Carlip \cite{Carlip:2000nv} and suggests that the additional entropy added to the black hole when the probe is absorbed can be computed using Cardy's formula. 


\acknowledgments
{\em Acknowledgements:} 
D. B. would like to thank T. Banks, J. Maldacena, D. Marolf, J. Polchinski, E. Silverstein for discussions.
Work supported in part by DOE under grant DE-FG02-91ER40618.
E. D. is supported by the Department of Energy Office of Science Graduate Fellowship Program (DOE SCGF), made possible in part by the American Recovery and Reinvestment Act of 2009, administered by ORISE-ORAU under contract no. DE-AC05-06OR23100.


\begin{thebibliography}{99}
\bibitem{Hawking:1974sw} 
  S.~W.~Hawking,
  Commun.\ Math.\ Phys.\  {\bf 43}, 199 (1975)
  [Erratum-ibid.\  {\bf 46}, 206 (1976)].


\bibitem{Braunstein:2009my} 
S. L. Braunstein, arXiv:0907.1190v1 (2009), published as 
  S.~L.~Braunstein, S.~Pirandola and K.~$\dot Z$yczkowski,
  Physical Review Letters 110, {\bf 101301} (2013)
  [arXiv:0907.1190 [quant-ph]].

\bibitem{Almheiri:2012rt} 
  A.~Almheiri, D.~Marolf, J.~Polchinski and J.~Sully,
  JHEP {\bf 1302}, 062 (2013)
  [arXiv:1207.3123 [hep-th]].


\bibitem{Almheiri:2013hfa} 
  A.~Almheiri, D.~Marolf, J.~Polchinski, D.~Stanford and J.~Sully,
  JHEP {\bf 1309}, 018 (2013)
  [arXiv:1304.6483 [hep-th]].


\bibitem{Marolf:2013dba} 
  D.~Marolf and J.~Polchinski,
  Phys.\ Rev.\ Lett.\  {\bf 111}, 171301 (2013)
  [arXiv:1307.4706 [hep-th]].


\bibitem{Bousso:2013wia} 
  R.~Bousso,
  Phys.\ Rev.\ D {\bf 88}, 084035 (2013)
  [arXiv:1308.2665 [hep-th]].


\bibitem{Maldacena:1997re} 
  J.~M.~Maldacena,
  Adv.\ Theor.\ Math.\ Phys.\  {\bf 2}, 231 (1998)
  [hep-th/9711200].


\bibitem{Banks:1996vh} 
  T.~Banks, W.~Fischler, S.~H.~Shenker and L.~Susskind,
  Phys.\ Rev.\ D {\bf 55}, 5112 (1997)
  [hep-th/9610043].


\bibitem{Asplund:2011qj} 
  C.~Asplund, D.~Berenstein and D.~Trancanelli,
  Phys.\ Rev.\ Lett.\  {\bf 107}, 171602 (2011)
  [arXiv:1104.5469 [hep-th]].


\bibitem{Asplund:2012tg} 
  C.~T.~Asplund, D.~Berenstein and E.~Dzienkowski,
  Phys.\ Rev.\ D {\bf 87}, 084044 (2013)
  [arXiv:1211.3425 [hep-th]].


\bibitem{Berenstein:2012ts} 
  D.~Berenstein and E.~Dzienkowski,
  Phys.\ Rev.\ D {\bf 86}, 086001 (2012)
  [arXiv:1204.2788 [hep-th]].


\bibitem{Iizuka:2001cw} 
  N.~Iizuka, D.~N.~Kabat, G.~Lifschytz and D.~A.~Lowe,
  Phys.\ Rev.\ D {\bf 65}, 024012 (2002)
  [hep-th/0108006].

\bibitem{Hanany:1996ie} 
  A.~Hanany and E.~Witten,
  Nucl.\ Phys.\ B {\bf 492}, 152 (1997)
  [hep-th/9611230].


\bibitem{Iizuka:2013yla} 
  N.~Iizuka, D.~Kabat, S.~Roy and D.~Sarkar,
  Phys.\ Rev.\ D {\bf 87}, 126010 (2013)
  [arXiv:1303.7278 [hep-th]].


\bibitem{Wiseman:2013cda} 
  T.~Wiseman,
  JHEP {\bf 1307}, 101 (2013)
  [arXiv:1304.3938 [hep-th]].


\bibitem{Anagnostopoulos:2007fw} 
  K.~N.~Anagnostopoulos, M.~Hanada, J.~Nishimura and S.~Takeuchi,
  Phys.\ Rev.\ Lett.\  {\bf 100}, 021601 (2008)
  [arXiv:0707.4454 [hep-th]].


\bibitem{Catterall:2008yz} 
  S.~Catterall and T.~Wiseman,
  Phys.\ Rev.\ D {\bf 78}, 041502 (2008)
  [arXiv:0803.4273 [hep-th]].


\bibitem{Hanada:2008ez} 
  M.~Hanada, Y.~Hyakutake, J.~Nishimura and S.~Takeuchi,
  Phys.\ Rev.\ Lett.\  {\bf 102}, 191602 (2009)
  [arXiv:0811.3102 [hep-th]].


\bibitem{Kaplunovsky:1983ev} 
  V.~Kaplunovsky and M.~Weinstein,
  Phys.\ Rev.\ D {\bf 31}, 1879 (1985).


\bibitem{Kabat:1998vc} 
  D.~N.~Kabat and G.~Lifschytz,
  JHEP {\bf 9812}, 002 (1998)
  [hep-th/9806214].


\bibitem{Susskind:1993ki} 
  L.~Susskind,
  Phys.\ Rev.\ Lett.\  {\bf 71}, 2367 (1993)
  [hep-th/9307168].


\bibitem{Taylor:1998tv} 
  W.~Taylor and M.~Van Raamsdonk,
  JHEP {\bf 9904}, 013 (1999)
  [hep-th/9812239].


\bibitem{Taylor:1999gq} 
  W.~Taylor and M.~Van Raamsdonk,
  Nucl.\ Phys.\ B {\bf 558}, 63 (1999)
  [hep-th/9904095].


\bibitem{Horowitz:2009wm} 
  G.~Horowitz, A.~Lawrence and E.~Silverstein,
  JHEP {\bf 0907}, 057 (2009)
  [arXiv:0904.3922 [hep-th]].


\bibitem{Ambjorn:2005db} 
  J.~Ambjorn, J.~Jurkiewicz and R.~Loll,
  Phys.\ Rev.\ Lett.\  {\bf 95}, 171301 (2005)
  [hep-th/0505113].


\bibitem{Carlip:2009km} 
  S.~Carlip,
  arXiv:1009.1136 [gr-qc].


\bibitem{Mathur:2005zp} 
  S.~D.~Mathur,
  Fortsch.\ Phys.\  {\bf 53}, 793 (2005)
  [hep-th/0502050].


\bibitem{Carlip:2000nv} 
  S.~Carlip,
  Class.\ Quant.\ Grav.\  {\bf 17}, 4175 (2000)
  [gr-qc/0005017].

\end{thebibliography}
\end{document}